\documentclass[aps,prb,superscriptaddress,nobalancelastpage,citeautoscript,amsmath,amssymb,floatfix,showkeys,natbib,twocolumn]{revtex4-2}
\usepackage[T1]{fontenc}
\usepackage[latin2]{inputenc}
\usepackage[dvipsnames]{xcolor}
\usepackage{graphicx}
\usepackage{soul}
\usepackage{dcolumn}
\usepackage{amsmath}
\usepackage{amssymb}
\usepackage[%
colorlinks=true,
urlcolor=blue,
linkcolor=blue,
citecolor=blue
]{hyperref}

\def  \bsig    {\mbox{\boldmath$\sigma$}}
\def  \bta     {\mbox{\boldmath$\tau$}}

\keywords{graphene, magnetic domain wall, magnetized graphene, Rashba spin-orbit interaction, localized states, edge states}

\begin{document}
\title{Graphene with Rashba spin-orbit interaction and coupling to a magnetic layer:  electron states localized at the domain wall}

\author{M. Inglot}
\affiliation{Department of Physics and Medical Engineering, Rzesz\'ow University of Technology,
al.~Powsta\'nc\'ow Warszawy 6, 35-959 Rzesz\'ow, Poland}

\author{V. K. Dugaev}
\affiliation{Department of Physics and Medical Engineering, Rzesz\'ow University of Technology,
al.~Powsta\'nc\'ow Warszawy 6, 35-959 Rzesz\'ow, Poland}

\author{A. Dyrda\l}
\affiliation{Faculty of Physics, Adam Mickiewicz University,
ul. Uniwersytetu Pozna\'nskiego 2, 61-614 Pozna\'n, Poland}

\author{J. Barna\'s}
\affiliation{Faculty of Physics, Adam Mickiewicz University,
ul. Uniwersytetu Pozna\'nskiego 2, 61-614 Pozna\'n, Poland}
\affiliation{Institute of Molecular Physics, Polish Academy of Sciences, ul. M. Smoluchowskiego 17,
60-179 Pozna\'n, Poland}

\begin{abstract}
Electron states localized  at a magnetic domain wall in a  graphene caplayer with Rashba spin-orbit interaction and coupled to a magnetic overlayer are studied theoretically. It is shown that two one-dimensional  bands of edge modes propagating along the domain wall emerge in the energy gap for each Dirac point, and the modes associated with different Dirac points K and K' are the same.
The coefficients describing decay of the corresponding wavefunctions  with distance from the domain wall contain generally real and imaginary terms.
Numerical results
on the local spin density and on the total spin expected in the edge states characterized by the wavenumber $k_y$ are presented and discussed.
The Chern number for a single magnetic domain on graphene indicates that the system is in the quantum anomalous Hall phase, with two chiral modes at the edges.  In turn, the number of modes localized at the domain wall is determined  by  the difference in Chern numbers on both sides of the wall. These numbers are equal to 2 and -2, respectively, so there are four modes localized at the domain wall.
\end{abstract}

\date{\today}
\maketitle

\section{Introduction \label{secIntr}}

It is well known that a two-dimensional electron gas  appears at the interface of two different insulators with nonequivalent topology of electron bands \cite{Hasan2010,Qi2011}. A typical example is the interface between an ordinary insulator (or vacuum) and a three-dimensional topological insulator (for instance Bi$_2$Te$_3$) \cite{Zhang2009} or a crystalline topological insulator~\cite{Fu2011,Ando2015}. The low-energy spectrum of electron states at the interface can be then  described by the massless relativistic Dirac Hamiltonian.

A characteristic feature of the topological non-equivalence of two materials in contact is the inversion of energy bands at the interface. An interesting example is the 2D Dirac electron gas with perpendicular magnetization that induces the energy gap $\Delta $ in the Dirac spectrum \cite{Liu2009a,Chen2010,Ferreiros2015}. This spectrum does not depend on the sign of $\Delta $, however the energy bands become inverted at the interface between regions with  $\Delta >0$ and $\Delta <0$. As a result, an additional one-dimensional energy band  of electron states
localized at the boundary  separating the areas of $\Delta >0$ and $\Delta <0$
appears in the system. This can be also considered as the appearance of electron states coupled to the magnetic domain wall. Interestingly, such electron states in topological insulators with a magnetic layer on top are responsible for nondissipative equilibrium currents along the domain wall \cite{Yasuda2017,Sedlmayr2020a,Araki2016}.

It should be noted that the basic idea of electron states bound to the kink of a static scalar field was formulated long ago by Jackiw and Rebbi~\cite{Jackiw1976}, who demonstrated the existence of zero-energy electron states in the  systems of Dirac and Yang-Mills fermions. Using various realizations of this idea one can find zero-energy solutions at the contact of narrow-gap semiconductors with mutually inverted energy bands \cite{Volkov1986}, at the vortices in chiral superconductors \cite{Volovik1999}, at hedgehogs in superconductors with coexisting singlet and triplet pairing \cite{Nishida2010}, and in the spectrum of surface electrons with a gap inversion in topological insulators \cite{Lee2007}.

It has been shown recently that the spin-orbit interaction can play an important role when considering the edge states,  leading e.g. to  spin polarization of the boundary. An example is a sharp p-n junction in graphene in the presence of spin-orbit coupling and magnetic field \cite{Bercioux2019}. In such a case electron zero modes with linear dispersion appear at the p-n junction, and the corresponding electron states are spin polarized.
In one-dimensional models with Rashba spin-orbit coupling (Rashba nanowires) and external magnetic field, some  unusual properties (e.g., equilibrium spin currents and localized spin torque) can appear, which are related to emerging edge states at the boundaries between magnetic \cite{Ronetti2020} or Rashba-coupling \cite{Dolcini2018,Gani2020} domain walls.

In this paper we consider a graphene-based structure consisting of a graphene monolayer deposited on a substrate that ensures the Rashba spin-orbit interaction~\cite{Dolcini2017,Ju2015} and covered by a magnetic layer with a domain wall, as presented in Fig.~\ref{fig1}~(a). The magnetic and spin orbital proximity effects induced in graphene are important ingredients of the model, since both of them
modify the energy spectrum substantially.  The magnetization of capping layer is assumed to be perpendicular to the graphene plane (i.e., it is along the axis $z$ in Fig.~\ref{fig1}) and coupled to the graphene either by exchange or stray fields.
A uniform proximity-induced magnetization in graphene (no spin-orbit coupling) shifts the spin up/down bands   upward/downward respectively, but leaves the two zero-energy crossing points in the vicinity of the K and K' points, see Fig.~\ref{fig1}~(b,c). In turn, the Rashba spin-orbit interaction induces spin-mixing and lifts the four-fold degeneracy at the K/K' points, as presented in Fig.~\ref{fig1}~(d)~\cite{Escudero2017}.
 When both proximity-induced magnetization and Rashba spin-orbit interaction are present in the system, the bulk energy gap is opened and all the four bands around Dirac points are non-degenerate, see Fig.~\ref{fig1}~(e,f). Thus, both magnetic and spin-orbit proximity effects enable controlling  the electronic structure and   also electric and magnetic properties of the graphene-based systems under consideration. Importantly, when Rashba spin-orbit interaction and magnetization (Zeeman-like field) are simultaneously present in graphene, one can observe the quantum anomalous Hall effect phase with quantized value of the Hall conductance when the Fermi level is inside the energy gap.

We show that creating a domain wall in the magnetic layer leads to further possibilities of controlling electronic and transport properties \cite{Ono1999,Truetzschler2016,Parkin2008}. In particular, we show that the domain wall generates conductive states inside the bulk energy gap. These states are localized at the domain wall and lead to  additional functionality of the graphene-based structure because the magnetic domain wall  can be controlled by current and/or magnetic field \cite{Saitoh2004,Allwood2005,Hayashi2008,Krzysteczko2017}.

It is worth noting that the influence of domain walls on electronic spectrum in graphene has been already discussed in the literature. However the domain walls were of different origin and were related to the possible stacking faults in  bilayer \cite{Ju2015} or multilayer \cite{Lee2016} graphene. Another type of  domain walls in gaped graphene, which can appear due to a substrate (like hexagonal Boron Nitride)  with a linear symmetry-breaking defect, was  considered by Semenoff et al. \cite{Semenoff2008}. They demonstrated the existence of localized states at the domain walls, and pointed out their importance  for possible applications.

Using the effective model describing low-energy excitations in a magnetized graphene with Rashba spin-orbit interaction, we calculate the energy and wave functions of the edge states localized at the magnetic domain wall. These  modes exist in the gap and propagate along the domain wall. Furthermore, the modes with opposite wave vectors (with respect to K/K' point) have  different energy, $\varepsilon (k_y)\ne \varepsilon (-k_y)$.
We show that two edge modes appear in the spectrum for each Dirac point, and the modes associated with different Dirac points K and K' are the same.
We also show that the attenuation factors that describe decay of the wavefunctions of chiral modes with distance from the domain wall contain imaginary terms. Accordingly, the corresponding local values of expected physical quantities include an oscillatory contribution with the  amplitude decaying with the distance from the wall.
As an example, we present numerical results
on the local spin density and total spin expected in the edge states.

In section \ref{secII} we describe the model studied in this paper. Electronic states localized at the domain wall are calculated in section \ref{secIII} for $k_y=0$, and in Sec. \ref{secIV} for the case of nonzero $k_y$. In Sec. \ref{secIV} we  also present dispersion curves of the modes propagating along the wall for both  K and K' Dirac points. Numerical results on the local spin density in the edge states are presented and discussed in Sec. \ref{secV}. Topological aspects are studied in Sec. \ref{secVI}, whereas summary and final conclusions  are in Sec. \ref{secVIII}. Symmetry relations of the model and scattering processes are discussed in the Appendices A and B, respectively .

\section{Model of graphene with a magnetic domain wall \label{secII}}

\begin{figure}
\includegraphics[width=\columnwidth]{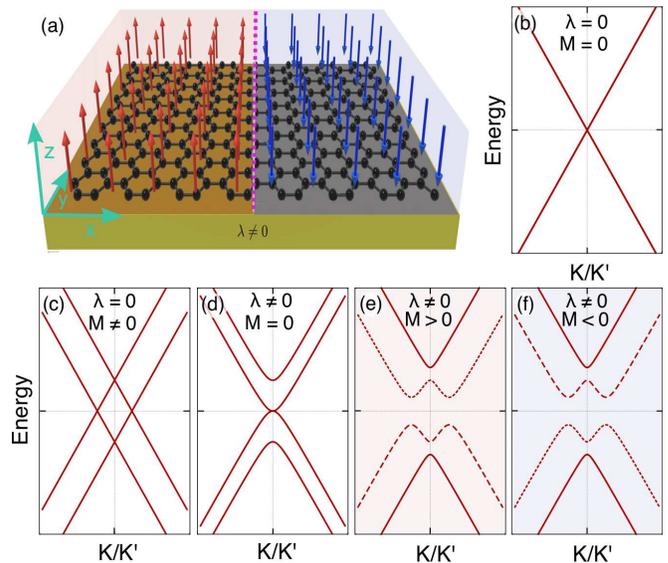}
\caption{
(a) Schematic picture of the structure: single layer of graphene sandwiched between a substrate inducing Rashba spin-orbit coupling and a magnetic layer with a domain wall. (b-f) Schematic band structure of graphene in the absence of magnetization and Rashba coupling (b);  in the presence of magnetization, $M\neq 0$ (c); in the presence of Rashba coupling, $\lambda \neq 0$ (d); and  in the presence of both magnetization and Rashba couling (e,f). Though the band structure for $M>0$ and $M<0$ is the same, the bands on the two sides are inverted.
}
\label{fig1}
\end{figure}

We consider a graphene monolayer deposited on a substrate which generates Rashba spin-orbit interaction\cite{Wang2016,Frank2016,Yang2016,Klimovskikh2015,Leicht2014}. In addition, we assume  a thin magnetic layer on top of the graphene~\cite{Khodas2009} with magnetization perpendicular to the graphene plane. Coupling to the magnetization opens then a gap in the electronic spectrum of graphene. In this paper we consider a more general situation, when the magnetization is not uniform but forms two domains  separated by a narrow domain wall as shown schematically in Fig.~\ref{fig1}~(a).

Effective Hamiltonian describing low-energy electronic states near the K point of the Brillouinn zone in the system under consideration can be written in the form \cite{Kane2005}
\begin{eqnarray}
\label{1}
\hat{H}_K =-iv\, (\tau _x\partial _x+\tau _y\partial _y) + \lambda(\sigma_y\tau_x-\sigma_x\tau_y) + \sigma _z M(x) ,\hskip0.5cm
\end{eqnarray}
where $v=\hbar v_F\simeq \hbar c/300$, $\bta$ and $\bsig$ represent the vectors of
Pauli matrices in the sublattice and spin spaces, respectively, $\lambda$
is the Rashba spin-orbit coupling parameter,
and $M(x)$ is the $x$-dependent gap parameter that is related to the magnetization in the $z$-direction (perpendicular to the graphene plane).
We assume  $M(x)$   in the following form \cite{Semenoff2008}:
\begin{align}
M(x)=\left\{ \begin{array}{cc}
M_0, & x<0, \\ -M_0, & x\geqslant 0,
\end{array} \right.
\end{align}
which describes a sharp magnetic domain wall located at $x=0$ and uniform along the $y$-axis, as shown schematically in Fig.~\ref{fig1}~(a). We note, that such very sharp domain walls can be created artificially in real systems \cite{Frackowiak2020}.
The bulk (two-dimensional) electronic band structure  of graphene corresponding to  $M>0$ is the same as that for $M<0$, as presented in Fig.~\ref{fig1}~(e) and (f), respectively. However, the bands become inverted when $M$ changes sign at the domain wall, which is also indicated in  Fig.~\ref{fig1}(e,f).

The energy gap in the  electronic spectrum of a uniformly magnetized graphene (no domain wall) with Rashba spin-orbit interaction is given by the  following formula \cite{Dyrdal2017}:
\begin{align}
E_g = \frac{2|M_0 \lambda|}{\sqrt{M_0^2 + \lambda^2}}.
\end{align}
This gap is determined by the absolute value of the magnetization, $|M|=M_0$, and absolute value of the Rashba parameter, $|\lambda |$. Note,  the gap vanishes when either $M_0=0$ or $\lambda = 0$.

Due to the band inversion, electronic states localized at the domain wall emerge in the energy gap.
Using the Schr\"odinger equation, $(\hat H -\varepsilon)\, \psi({\bf r})=0$, and taking into account structure geometry, one can write the wave function in the form $\psi({\bf r}) = e^{ik_y y}\, \psi_{k_y}(x)$, where $\psi_{k_y}(x)$ is a bispinor with four components,
\begin{align}
\psi_{k_y}^T =
\left(
\varphi _{k_y}^\uparrow, \varphi _{k_y}^\downarrow, \chi_{k_y}^\uparrow, \chi_{k_y}^\downarrow
\right) .
\end{align}
In the following we will solve the Schr\"odinger equation and calculate the energy spectrum of the edge states localized at the  domain wall.

\section{States localized at the domain wall for $k_y=0$\label{secIII}}

Let us consider first the states localized at the domain wall for  $k_y=0$. Equations for the wave function components (4) for $x<0$ and $x>0$ acquire then the following form:
\begin{align}
\label{5}
&(M-\varepsilon)\, \varphi_{0}^{\uparrow}-iv\, \partial_x \chi_{0}^\uparrow = 0,
\nonumber\\
&(M+\varepsilon)\, \varphi_{0}^{\downarrow} - 2i \lambda \chi_{0}^{\uparrow} +i v\, \partial_x \varphi_{0}^{\downarrow} = 0,
\nonumber\\
&iv\, \partial_x  \varphi_{0}^{\uparrow} + 2i \lambda \varphi_{0}^{\downarrow} - (M-\varepsilon)\, \chi_{0}^{\uparrow} = 0,
\nonumber\\
&iv\, \partial_x \varphi_{0}^{\downarrow} + (M + \varepsilon)\, \chi_{0}^{\downarrow} = 0,
\end{align}
where $M=M_0$ ($x<0$) and $M=-M_0$ ($x>0$).
For states localized at the domain wall one can write the  wavefunctions $\chi_{0}^{\uparrow ,\downarrow}$ and $\varphi_{0}^{\uparrow ,\downarrow}$ in the form
\begin{eqnarray}
\label{content...}
\varphi_0^\uparrow (x)= Ae^{\kappa x} ,\hskip0.5cm
\varphi_{0}^\downarrow (x)=B e^{\kappa x},
\nonumber \\
\chi_{0}^\uparrow (x)= Ce^{\kappa x}, \hskip0.5cm
\chi_{0}^\downarrow (x)=D e^{\kappa x},
\end{eqnarray}
where $A,B,C,D$ are certain constants, while $\kappa$ describes a wavefunction decay on both sides of the domain wall. Thus, real value of $\kappa$ must be positive,  ${\rm Re}\, \kappa > 0$, for $x<0$ and negative, ${\rm Re}\, \kappa < 0$, for $x>0$.  Upon substituting Eq.(6) into Eq.(5) one obtains a system of linear algebraic equations for the constants $A,B,C,D$, which has nonzero solutions if the corresponding determinant vanishes. This leads to the following equation for $\kappa $:
\begin{align}
\label{7}
\kappa ^4 + 2 M_0^2 \kappa ^2 + 2 \varepsilon^2 \kappa^2 + M_0^4 - 2 M_0^2 \varepsilon^2 \\ \nonumber
+ 4 M_0^2 \lambda^2  - 4 \varepsilon^2 \lambda^2 + \varepsilon^4 = 0.
\end{align}
Note, this equation holds for $x>0$ and $x<0$.
From this we find four solutions denoted as $\kappa_n$ ($n=1,3$) and $\kappa_p$ ($p=2,4$),
\begin{align}
\label{8}
\kappa_{n} & = \frac1{v}
\Big( - M_0^2 - \varepsilon^2 \pm 2 \sqrt{M_0^2 \varepsilon^2 - M_0^2 \lambda^2 + \varepsilon^2 \lambda^2}\Big) ^{1/2} ,
\\
\kappa_{p} & = - \frac1{v} \Big(- M_0^2 - \varepsilon^2 \pm 2 \sqrt{M_0^2 \varepsilon^2 - M_0^2 \lambda^2 + \varepsilon^2 \lambda^2}\Big) ^{1/2}.
\label{8b}
\end{align}
The above solutions for $\kappa$ are complex in general.
However we find that ${\rm Re}\, \kappa _1={\rm Re}\, \kappa _3>0$ (so they correspond to $x<0$)   and ${\rm Re}\, \kappa _2={\rm Re}\, \kappa _4<0$ (and correspond to $x>0$).

\begin{figure}
    \centering
	\includegraphics[width=0.99\columnwidth]{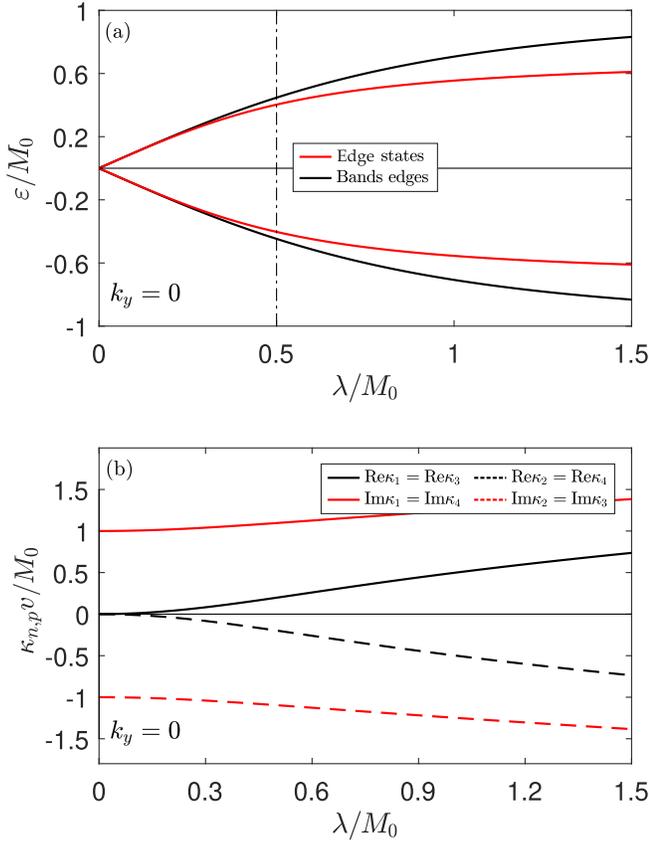}
    \caption{(a) Energy  of the electron states (for $k_y=0$) localized at the domain wall (red lines) as a function of  $\lambda$  for $M_0=20$~meV. Black lines correspond to the edges of the gap in the bulk spectrum. Both, the energy $\epsilon$ and Rashba parameter $\lambda$ are normalized to $M_0$. The vertical dashed-dot line marks the value of $\lambda/M_0$ used in the following figures. (b) Real and imaginary parts of $\kappa_{n,p}$ (normalized to $M_0/v$) defined by Eqs.~(8,9) for $k_y=0$.
    }
    \label{fig2}
\end{figure}

Upon determining the coefficients $A,B,C,D$ in Eqs.~(6), one can write  two possible solutions of the Schr\"odinger equation
for $x<0$ and $x>0$ in the following form,
\begin{align}
	\label{9}
	&\psi_{n(p)}(x)=\nonumber\\
	&e^{\kappa _{n(p)}x}
	\left(
	\begin{array}{c}
		1\\
		\displaystyle{-\frac{ -k_y^{2} v^{2}+\kappa_{n(p)}^{2} v^{2} + (M\pm\varepsilon)^{2}}{2 \lambda  v \left(k_y +\kappa_{n(p)}\right)}} \\[10pt]
		\pm\displaystyle{\frac{i \, \left(M \pm\varepsilon \right)}{v \left(k_y +\kappa_{n(p)}\right)}}\\[10pt]
		\pm\displaystyle{\frac{i  \left(k_y -\kappa_{n(p)}\right) \left(\kappa_{n(p)}^{2} v^{2}-k_y^{2} v^{2}+\left(\varepsilon +M \right)^{2}\right)}{2\lambda  \left(k_y +\kappa_{n(p)}\right) \left( M \mp\varepsilon\right)}}
	\end{array}
	\right),
\end{align}
where the index $n$ and the upper sign correspond to $x<0$, whereas the index $p$  and lower sign correspond to $x>0$.
Since the real part of $\kappa _n$ is positive and that of $\kappa _p$ is negative, one can write a general normalized  wave function corresponding to the states localized at the domain wall, i.e. the wave function that exponentially decays with distance from the wall on both sides (for $x<0$ and $x>0$)  in the form
\begin{eqnarray}
\label{}
\psi (x)=N[a_1\psi _1(x)+a_3\psi _3(x)],\hskip0.5cm x<0,
\\
\psi (x)=N[a_2\psi _2(x)+a_4\psi _4(x)],\hskip0.5cm x>0,
\end{eqnarray}
where $a_1,...a_4$ are certain coefficients, which have to be determined from the continuity condition of the wave function at $x=0$, and $N$ is a normalization factor.
From this condition one obtains a system of four linear algebraic equations for $a_1,...a_4$ in Eqs.~(11) and (12). Vanishing of the corresponding determinant defines energy of the localized states, which can be formally presented in the simple analytical form
\begin{align}
	\varepsilon_{1,2} &= \frac13\, \sqrt{3\,{M_0}^{2}+4\,{\lambda}^{2}}
	\left( \sqrt {3} \sin \gamma_{1,2} - \cos \gamma_{1,2}\right) \pm \frac{2\lambda}{3},
\end{align}
where
\begin{align}
	\gamma_{1} &= \left\{
	\begin{array}{ll}
		\frac 13\,\arctan \left({\frac {12\sqrt {3}\left| M_0\right| \xi_1 }{\xi_2 }}  \right) + \frac{\pi}{3}, & \; \xi_2 <0\\
		\frac 13\,\arctan \left({\frac {12\sqrt {3}\left| M_0\right| \xi_1 }{\xi_2 }}  \right) , & \; \xi_2 >0
	\end{array}
	\right.
	\\
	\gamma_{2} &= \left\{
	\begin{array}{ll}
		\frac 13\,\arctan \left({\frac {3\sqrt {3}\left| M_0\right| \xi_1 }{-\xi_2 }} \right),& \; \xi_2 <0\\
		\frac 13\,\arctan \left({\frac {3\sqrt {3}\left| M_0\right| \xi_1 }{-\xi_2 }} \right) + \frac{\pi}{3},  & \; \xi_2 >0
	\end{array}
	\right. ,
	\label{Eq11}
\end{align}
with $\xi_1 = \sqrt {4\,{M_0}^{4}+13\,{M_0}^{2}{\lambda}^{2}+32\,{\lambda}^{4}}$ and
$\xi_2 = -36\,{M_0}^{2}\lambda+64\,{\lambda}^{3}$.

\begin{figure}
    \centering
   \includegraphics[width=0.91\columnwidth]{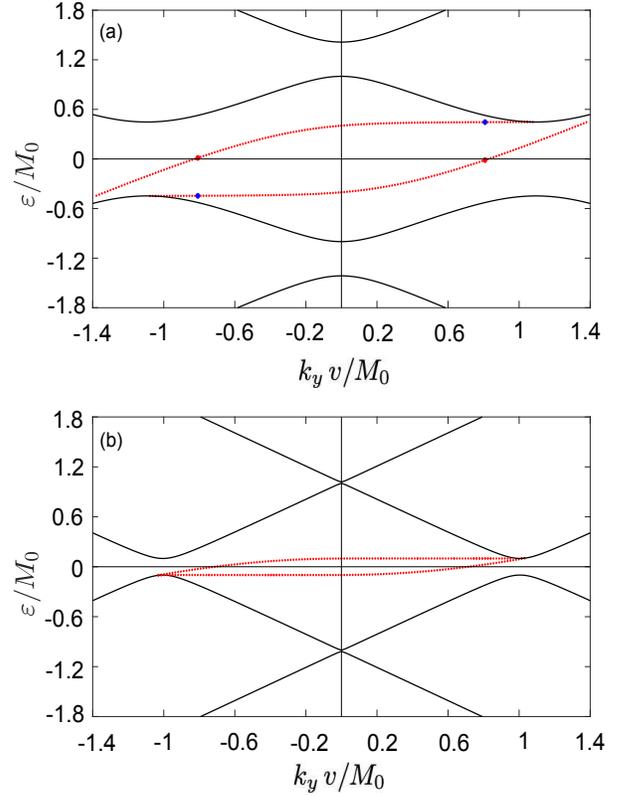}
     \caption{Edge modes (red dot lines) localized at the domain wall, presented  as a function of $k_y$ (normalized to $M_0/v$)
     for (a) $\lambda/M_0=0.5$,  $\lambda= 10$~meV,  and (b) $\lambda/M_{0} = 0.1$ ,  $\lambda = 2$~meV. In both plots $M_0=20$~meV.  The solid black lines determine the relevant conduction and valance bulk bands of graphene with Rashba spin-orbit coupling and magnetization. The large red and blue  dots on the dispersion curves of the edge states correspond to the modes chosen in Figs.~\ref{fig4} and \ref{fig5}. The two red points (as well as the blue ones) are related by symmetry. Dispersion curves of the edge modes in the K and K' points are the same.}
    \label{fig3}
\end{figure}

The localized states described by Eq.(13) exist inside the energy gap. This is shown explicitly in Fig.~\ref{fig2}(a)  where the two energy levels $\varepsilon _{1,2}$ are presented  by the red lines as a function of the spin-orbit coupling constant $\lambda $ normalized to $M_0$.
 We remind that these energy levels  correspond to  $k_y=0$.
The black lines in this figure describe the valence and conduction band edges (and thus determine the energy gap). The corresponding real and imaginary parts of the parameters $\kappa_{n,p}$ (normalized to $M_0/v$) are shown in Fig.~\ref{fig2}(b).
This figure clearly shows that $\kappa_{1,3}$ have positive real parts, and thus describe exponential localization  of the wavefinction on the left side ($x<0$), while the real parts of $\kappa_{2,4}$ are negative and  describe exponential localization at the domain wall on the right side of the  wall ($x>0$). All the parameters $\kappa$ have also imaginary parts.

\section{Dispersion curves  of the edge states\label{secIV}}

Now we determine the modes localized at the domain wall for nonzero values of $k_y$.
All the calculation steps for $k_y\ne 0$ are similar to those for $k_y=0$, but the derived formulae are cumbersome so they  will not be presented here. Instead, we will show  some numerical results.  Energy $\varepsilon _{1(2)}$  of the lower (higher) edge mode is presented in Fig.~\ref{fig3}(a) as a function of $k_y$ normalized to $M_0/v$ ($k_y/(M_0/v) = k_yv/M_0$)
and for  $\lambda /M_0 =0.5$. These modes occur in the energy gap,
but when they enter the conduction or valence bands, they acquire quasi-localized (or resonant) character
due  to interaction with the bulk electron bands.
From symmetry (see Appendix A) follows that the whole spectrum is antisymmetric. This is clearly seen in Fig.~\ref{fig3}(a), where $\varepsilon_1(k_y)=
-\varepsilon_2(-k_y)$.

When the Rashba parameter $\lambda$ decreases, the two localized modes in each Dirac point close up, as shown in Fig.3(b) for a small value of the parameter $\lambda$. When $\lambda$  tends to zero, these modes become degenerate and their energy tends to zero. In addition, they acquire the bulk character in the limit $\lambda \rightarrow 0$ since the inverse localization length $ {\rm Re}\, \kappa$ tends then to zero, see also Fig.2(b).

\begin{figure}
    \centering
    \includegraphics[width=0.97\columnwidth]{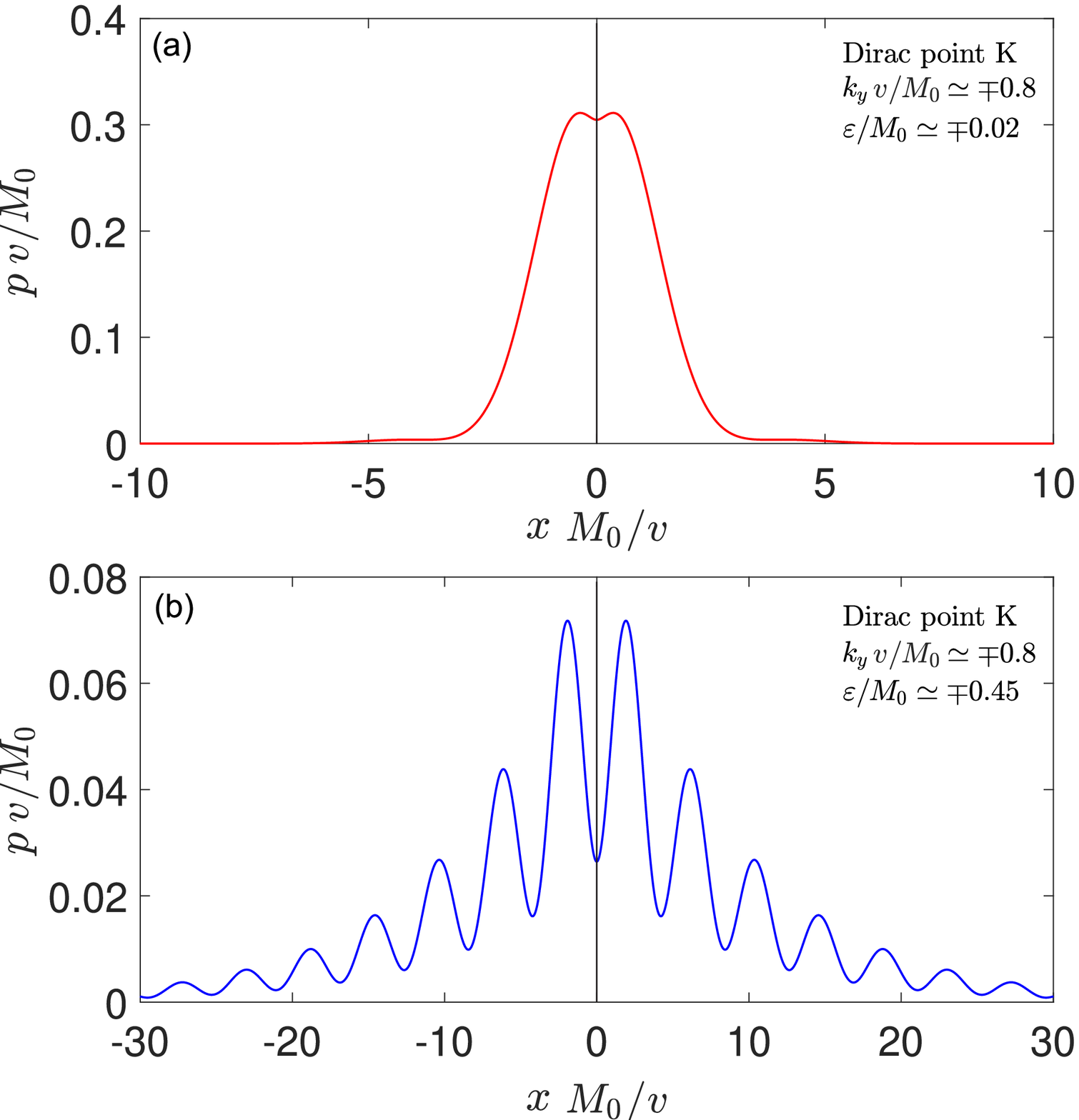}
    \caption{ (a) Normalized probability density $p$ as a function of $x$  for the edge states corresponding to the red (a) and blue (b) points on the dispersion curves in Fig.\ref{fig3}(a), and described  by the energy $\varepsilon /M$ and normalized wavevector $k_y$ as indicated.
Due to symmetry relations,  these curves describe the probability density for the $K$ and $K'$ Dirac points. Other parameters: $\lambda/M=0.5$ and $M_0 = 20$~meV.
}
\label{fig4}
\vspace{0.5cm}
%
    \centering
    \includegraphics[width=0.97\columnwidth]{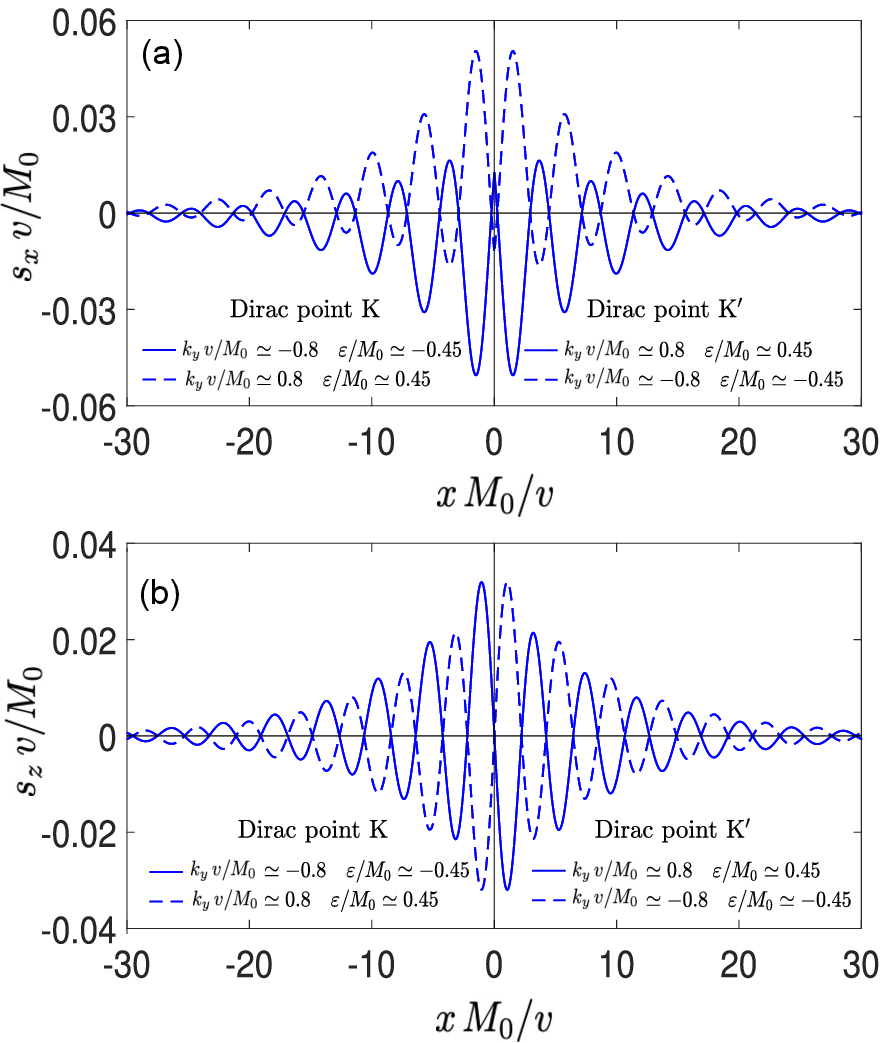}
    \caption{Normalized spin density $s_x$ (a) and $s_z$ (b) for the Dirac points $K$.
    The assumed parameters as indicated (they correspond to the blue points in Fig.3).
    Other parameters: $\lambda/M=0.5$ and $M_0 = 20$~meV.}
    \label{fig5}
\end{figure}

Due to the imaginary terms in $\kappa_{n,p}$, the wavefunctions have a nonzero oscillatory contributions, as clearly visible in Fig.\ref{fig4}, where the probability density $p=\psi ^\dag (x) \psi (x)$ (normalized to $M_0/v$) is shown as a function of the dimensionless parameter $xM_0/v$ (position on the axis $x$ perpendicular to the wall,  normalized to $v/M_0$)
 for the modes indicated by the red   and blue   dots on the dispersion curves of the edge  modes  in Fig.~\ref{fig3}(a) for the point $K$.
 This probability density decays on both sides with the distance from the domain wall. However, this decay has exponential and oscillatory contributions. Amplitude of the oscillatory term  decreases with increasing distance from the domain wall as well. As a result, expected values of some physical quantities in the edge states may behave in a similar manner.
 Period of the spacial oscillations is determined by the imaginary part of $\kappa $ in Eqs. (8) and (9). For small $\epsilon < \lambda $ and small $\lambda $ we get a period $\sim v/M_0$.
It is evident that the wavefunctions become more extended and the oscillations are more pronounced when energy of the edge modes approaches one of the two gap edges.

We emphasize that all the results presented up to now correspond to the Dirac point $K$ in the Brillouin zone. Similar modes exist also in the second Dirac point, $K'$. The corresponding Hamiltonian for the $K'$ point reads
\begin{eqnarray}
\hat{H}_{K'} =-iv\, (\tau _x\partial _x-\tau _y\partial _y) + \lambda(\sigma_y\tau_x+\sigma_x\tau_y) + \sigma _z M(x) ,\hskip0.6cm
\end{eqnarray}
Calculations similar to those described above for the point $K$ show that the edge modes (localized at the domain wall) associated with the point $K'$  are exactly the same as the modes corresponding to the point $K$. Accordingly, the results shown in Figs. 2, 3 and 4 apply also to the point $K'$.


\section{Spin density associated with the chiral states\label{secV}}

\begin{figure*}
	\centering
	\includegraphics[width=1.92\columnwidth]{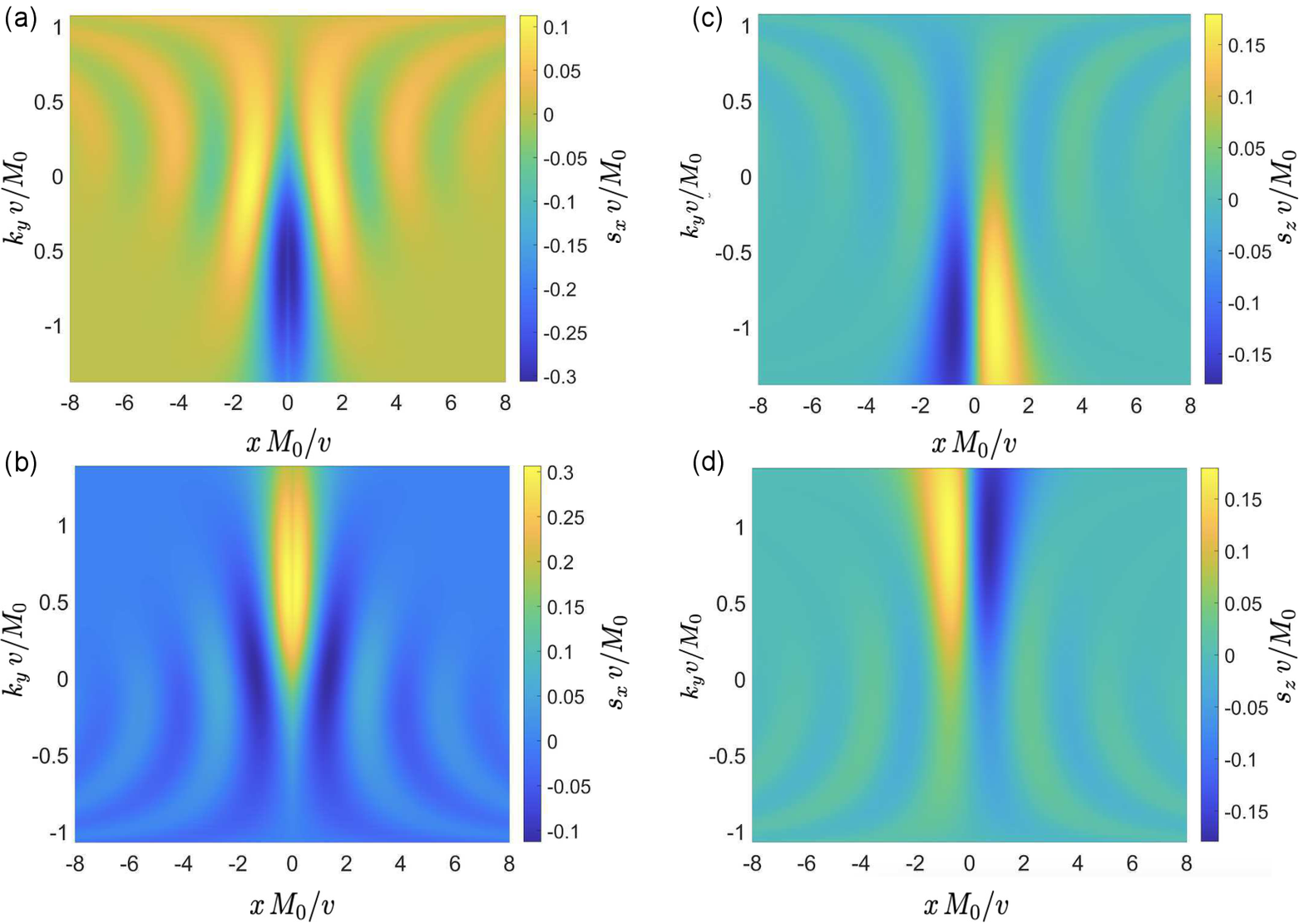}
	\caption{The $x$ and $z$ components of spin density associated with the edge mmodes of higher (a),(c) and lower (b), (d) energy, presented as a function of the corresponding normalized wavevector $k_y$ and normalized position $x$ on the axis normal to the domain wall. }
	\label{fig6nn}
\end{figure*}

Now we consider spatial variation of the spin density  associated with individual edge states, ${{\bf s}(x)=\psi ^\dag (x)\tau _0\, \bsig \psi (x)}$, where $\tau _0$ is the unit matrix in the sublattice space of graphene.
In Figs.~\ref{fig5}(a) we show  the  $x$  component of the spin density,  $s_x$, normalized to $M_0/v$ as a function of  $xM_0/v$ for the modes indicated by the blue points in Fig.\ref{fig3}(a) for the $K$ point. Note, the $x$-components of the spin density in these two symmetry-related points oscillates with increasing distance from the wall with opposite phases. Consequently, when both modes are populated, their contributions cancel each other. Apart from this, $s_x$ is a symmetric function of  $x$.
Qualitatively similar behavior can be observed for the $z$ components of the spin density, $s_z$. However, now the corresponding $x$ dependence is antisymmetric. In turn, the $y$-component of the spin density vanishes exactly, $s_y(x)=0$.

From Fig.5 follows that (i) the local spin density is oriented in the $(x,z)$ plane, and (ii) its  magnitude decays in an oscillatory manner with increasing distance from the domain wall. Oscillation period depends on the wavevector, as shown in Fig.~6, where the local density of $s_x$ and $s_z$ (normalized to $M_0/v$) is shown for both edge modes  as a function of the normalized wave vector and position on the axis $x$.


Here, it is interesting to note some similarity to the problem of {\it orthogonal spin polarization} at the Rashba-field domain wall in a homogeneously magnetized nanowire \cite{Rossi2020}. Even though the Hamiltonian of graphene differs substantially from the 1D Rashba model considered in Ref. [\onlinecite{Rossi2020}], a nonzero $s_x$ at the graphene domain wall corresponds to orthogonal spin polarization since the $x$-axis in our model is orthogonal to $M$ and to Rashba field when we consider electrons moving along the axis $x$ (like in the nanowire problem).

Let us analyse now the total expected spin components $S_x$ and $S_z$  in individual edge states considered above,  i.e. the corresponding spin  density integrated over $x$, $S_x=\int dx\, s_x(x)$ and $S_z=\int dx\, s_z(x)$. From the above analysis follows that $S_x$ in a particular chiral state is generally nonzero, while the $S_z$ component vanishes for all edge states. Therefore, we will  focus now only on the $S_x$ component.
For a particular energy there are two edge states in the gap. For each Dirac point we define the total spin $S^t_x(\varepsilon)$ associated with edge modes at this energy as a sum of the contributions from the two states. In Fig.7 we show $S^t_x$ as a function of energy. The total spin polarization of the edge states is opposite for the points $K$ and $K^\prime$. Thus, the integrated  spin polarization, obtained by integration over energy up to the Fermi level, vanishes due to compensation of the contributions from both Dirac points. To get a nonzero value of the integrated polarization one needs to lift the valley degeneracy.
\begin{figure}
	\centering
	\includegraphics[width=0.95\columnwidth]{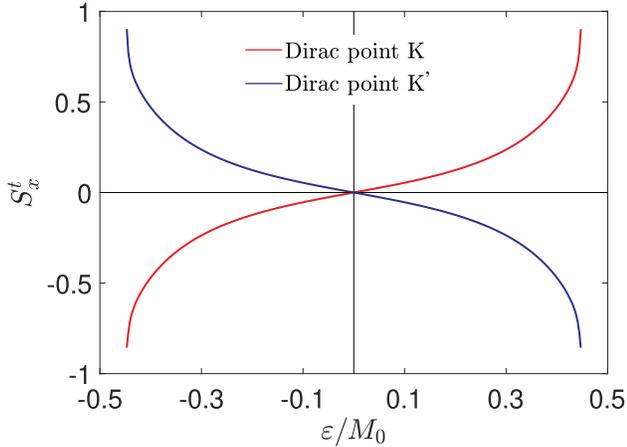}
	\caption{Total spin $S_x^t$ associated with the edge  states in the K and K' points, presented as a function of energy for ${\lambda/M=0.5}$.  }
	\label{fig7}
\end{figure}

\section{Bulk--edge correspondence and Quantum Anomalous Hall effect}
\label{secVI}

\begin{figure}
	\centering
	\includegraphics[width=\columnwidth]{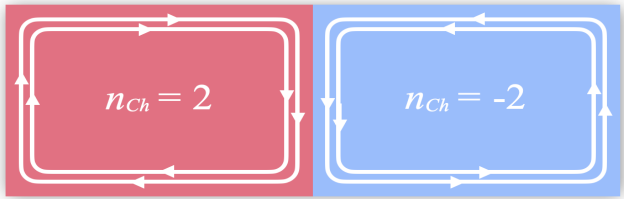}
	\caption{Schematic picture of the bulk--edge correspondence for the structure presented in Fig.~\ref{fig1}~(a). The graphene domains with opposite magnetizations possess opposite Chern numbers. Accordingly, the two domains possess counter propagating chiral edge states.}
	\label{fig9}
\end{figure}

The effective Hamiltonian describing bulk states on the right/left part of the system, i.e. of a single magnetic domain, is given by the extended Kane-Mele model \cite{Kane2005},
%
\begin{eqnarray}
\label{16}
\hat{H}_{\scriptscriptstyle{K,K'}}= v(\tau_x k_x +\eta_{z} \tau_y k_y)
+\lambda(\eta_{z} \sigma_x\tau_y+\sigma_y\tau_x)
+\sigma_z M , \nonumber\\
\end{eqnarray}
where we introduced the Pauli matrix $\eta_{z}$ acting in the valley subspace. Adding interaction with the perpendicular magnetization to the Hamiltonian of a pristine graphene leads to splitting of the energy bands, as presented in Fig.~\ref{fig1}~(c). In this case spin is still a good quantum number. The accidental crossing at two points with zero-energy can be easily removed by a spin-mixing term, that is by the Rashba spin-orbit interaction in our case. In turn, simultaneous action of  magnetization and Rashba field opens an energy gap in the bulk spectrum, see Fig.~\ref{fig1}~(e),(f).
Thus, the perpendicular to plane magnetization breaks the time-reversal symmetry, while the Rashba coupling is a consequence of the inversion symmetry breaking and leads to the mixing of spin states.
The insulating state, that appears when the energy gap is open, is topologically nontrivial and is called the Quantum Anomalous Hall (QAH) phase \cite{Haldane1988,Weng2015,Ren2016}. Accordingly,  when the Fermi energy is in the energy gap, one may expect the QAH conductance of the system. Based on the Thouless-Kohmoto-Nightingale-Nijs (TKNN) theory~\cite{Thouless1982,Niu1985}, the conductance is then given by the following simple expression:
\begin{equation}
\sigma_{xy} = \frac{e^{2}}{h} n_{\scriptscriptstyle{Ch}},
\end{equation}
where
\begin{equation}
n_{\scriptscriptstyle{Ch}} =  \frac{1}{2\pi} \sum_{\nu}  \sum_{n} \int d^{2}\mathbf{k}\; \Omega_{z,n}^{\nu}(\mathbf{k}) \equiv n_{\scriptscriptstyle{Ch}}^{K} + n_{\scriptscriptstyle{Ch}}^{K'}
\end{equation}
is the Chern number, and $\Omega_{z,n}^{\nu}(\mathbf{k})$ is the z-component of the Berry curvature for the n-th band in the momentum space ($\nu$ indicates one of the two inequivalent K points in graphene), whereas summation over $n$ is the summation over all occupied bands.
For the effective continuum model (Eq.~\ref{16}) describing bulk states in the system with a single magnetic domain we find  ${n_{\scriptscriptstyle{Ch}} = 2\,\mathrm{sgn}(M)}$, with the same contributions from the K and K' points, i.e. ${n_{\scriptscriptstyle{Ch}}^{K} = n_{\scriptscriptstyle{Ch}}^{K'} = 1\,\mathrm{sgn}(M)}$~\cite{Qiao2010,Qiao2012,Dyrdal2017,Hoegl2020}.  According to the bulk-boundary correspondence, one can then expect two chiral edge modes (see Fig.~\ref{fig9}),
at the interface between graphene and vaccum, that is between two topologically distinct phases, i.e. between the QAH phase and trivial insulator.

The  system with a magnetization kink (sharp domain wall) can be considered as a junction of two QAH insulators with opposite chiralities of the edge modes (the orientation of magnetization determines the sign of the Berry curvature) that have been connected adiabatically. This is reminiscent of the generalized Jackiw-Rebbi model for Dirac fermions in graphene with the magnetization kink, $M(x) = - M (-x)$. Accordingly we have a pair of topologically different topological insulators ($n_{\scriptscriptstyle{Ch}}(x < 0) = 2$ and $n_{\scriptscriptstyle{Ch}}(x > 0) = -2$), which  belong to the same symmetry class~\cite{Chiu2016}.
Thus, one can expect four edge states propagating in the same direction along the domain wall, as presented in Fig.~\ref{fig9}. The quantized anomalous Hall conductivity at the domain wall is equal to $-4 e^{2}/h$.

However,  it should be stressed that the edge states at the magnetization kink are not protected against scattering as the interface is not between topologically distinct insulators~\cite{Chiu2016,Cayssol2013}. More precisely, there are two topologically distinct regions (phases) defined by different  Chern numbers, but Hamiltonians describing these regions belong to the same class of topological order. Secondly, for graphene one cannot limit backward scattering processes to energy states around a single Dirac cone (single valley).
Note, that the time-reversal symmetry for graphene is defined by the operator $\Theta = i \eta_{x} \sigma_{y} \mathcal{K}$ \cite{Inoue2016a} where $\mathcal{K}$ is a complex conjugation. Due to the last term in the Hamiltonian (\ref{16}) the time-reversal symmetry is broken, i.e., ${\Theta H(\mathbf{k}) \Theta^{-1}  \neq H(-\mathbf{k})}$.
Consequently, the inter-valley scattering processes  can lead to localization of the edge modes. In Appendix~\ref{AppA} we provide a more detailed discussion of the symmetry of system under consideration.

The problem of intervalley scattering is an important issue in the context of the realization of QAH effect in real graphene based systems. It should be mentioned that the realization of graphene in the presence of the net perpendicular to plain magnetization and Rashba spin-orbit coupling is possible in several ways. One of the reported methods relies on the absorption of magnetic transition-metal adatoms  on one side of the graphene layer. This ensures not only magnetic proximity effect, but also a charge transfer between graphene and adatoms, which leads to a sizable Rashba effect \cite{Ding2011,Zhang2012,Dumeige2013,Lu2013a,Gong2015}. This solution is however related to the inter-valley scattering and diminishing of the QAH phase. Jing et al. \cite{Jiang2012} showed that the QAH phase can survive when the magnetic adatoms are distributed in a completely random way. Such a solution is rather difficult to achieve in realistic samples, as adatoms in graphene prefer to form clusters that destroy the QAH phase \cite{Eelbo2013,Chen2013}. Another realization is based on the deposition of graphene on ferromagnetic insulating substrate like YIG, RbMnCl$_3$ or LaMnO$_3$ \cite{Qiao2014,Wang2015,Zhang2015}. However, the Rashba spin-orbit coupling in such systems is weak and thus the energy gap is also rather small.
\section{Summary and conclusions}
 \label{secVIII}

We have considered energy spectrum of a system consisting of a graphene monolayer with spin-orbit Rashba interaction, which is additionally coupled to a magnetic layer with a domain wall. The main focus was on the  states localized at the domain wall, which exist in the gap created by coupling to the ferromagnetic layer in the presence of Rashba spin-orbit interaction. For each Dirac point we found two one-dimensional bands of chiral modes, and the modes in different points are similar. We have also calculated expected values of the spin density as well as the total spin expected in  the edge states. We have shown that the spin polarization of the edge states appears when there is an imbalance in the occupations of the  $K$ and $K'$ Dirac points.

From symmetry follows that $\varepsilon^K (k_y)=-\varepsilon^K (-k_y)$ (the same for the K' point), and also $\varepsilon^K (k_y)= \varepsilon^{K'} (k_y)$. Different domains correspond to  topologically different regions defined by different  Chern numbers (2 and -2 on different sides of the wall). From  this analysis follows that four chiral states propagate along the domain wall when the Fermi energy is in the gap, exactly what was found from direct calculations of the modes localized at the wall.  As a consequence, one can observe quantized anomalous Hall conductance along the domain wall with four conductance quanta.
However, the chiral states at the domain wall are not protected against inter-valley scattering. Such scattering processes can lead to localization of the edge modes and suppression of QAH effect. More details on scattering is given in Appendix B.

In our considerations we assumed a domain wall with equal magnitudes $|M|=M_0$ of magnetization on both sides.  When $M$ on one side of the wall is  small, the corresponding gap is narrow, and one can expect that two chiral modes at the wall are low-energy and slowly decay with distance from the wall. Obviously, when $M=0$ on one side, the second part of the system is then covered by a uniform magnetic domain and it is in the QAH phase, while the part with $M=0$ is in the metallic phase.


\begin{acknowledgements}
This work was partially supported by the National
Science Centre in Poland under the Project No. DEC-
2017/27/B/ST3/02881 (M.I., V.K., and J.B.) and by the Norwegian
Financial Mechanism under the Polish-Norwegian
Research Project NCN GRIEG "2Dtronics", Project
No. 2019/34/H/ST3/00515 (A.D.).
\end{acknowledgements}

\appendix
\section{Symmetry of the system}
\label{AppA}
Time-inversion operator for $\hat{H}(k_y)$ is
$T_t=\tau _y\sigma _y\mathcal{K}\mathcal{P}_{k_y}$,
where $\mathcal{K}$ stands for the complex conjugation and $\mathcal{P}_{k_y}$ changes $k_y\to -k_y$.
The Hamiltonian $\hat{H}_0$ is invariant with respect to this transformation,
$T_t\hat{H}_0T_t^\dag =\hat{H}_0$. The term $\sigma _zM$ in $\hat{H}$ breaks this symmetry, but reversing simultaniously the sign of $M$ leads to $T_t\, \hat{H}(M)\, T_t^\dag =\hat{H}(-M)$.

One can introduce electron-hole inversion operator $T_c=\tau _y\sigma _x\mathcal{K}$.
If $M$ is constant, we consider $\hat{H}_{K,K'}({\bf k})=v(\tau _xk_x\pm \tau _yk_y)+\lambda(\mp \sigma_x\tau_y+\sigma_y\tau_x)+\sigma _z M$.
In this case $T_c\, \hat{H}({\bf k})\, T_c^\dag =-\hat{H}({\bf k})$, which means that if $\psi _{\bf k}$ is the eigenfunction of $\hat{H}({\bf k})$ with the eigenvalue $\varepsilon ({\bf k})$, then $T_c\psi _{\bf k}$ is also the eigenfunction of $\hat{H}({\bf k})$ with the eigenvalue $-\varepsilon ({\bf k})$. This corresponds to electron-hole symmetry for $\hat{H}({\bf k})$.
When $M$ is not constant, eg. for $M(x)=-M(-x)$, this symmetry is broken.

Let us consider now the transformation $T_p=\tau _x$, for which we get
$T_p\, \hat{H}_K(k_y)\, T^\dag _p=\hat{H}_{K'}(k_y)$. This indicates that if $\psi _{k_y}(x)$ is the eigenfunction of $\hat{H}_K$ with the eigenvalue $\varepsilon (k_y)$ then the function $T_p\, \psi _{k_y}(x)$ is also the eigenfunction of $\hat{H}_{K'}$ with the same eigenvalue $\varepsilon (k_y)$. 

There is one more transformation $T_r=\sigma _z\mathcal{P}_x\mathcal{P}_{k_y}$,
which acts as $T_r\, \hat{H}_{K,K'}(k_y)\, T_r^\dag =-\hat{H}_{K,K'}(-k_y)$ when $M(x)=-M(-x)$ (this is the symmetry of our model). It leads to $\varepsilon (k_y)=-\varepsilon (-k_y)$. Thus, the dependence $\varepsilon (k_y)$ is an antisymmetric function of $k_y$.

Let us introduce vector ${\bf R}={\bf m}(-\delta )\times {\bf m}(+\delta )$, where ${\bf m}(x)={\bf M}(x)/M_0$. This vector determines the chirality. As one can conclude from Fig.~\ref{fig3}, the group velocity of electrons localized at the domain wall is in the direction of ${\bf R}$ for both valleys $K$ and  $K'$.

\section{Backward scattering from impurities\label{secVII}}

From the above symmetry considerations follows that for each Dirac point the $T_r$ symmetry relates the states $k_y$ and $-k_y$  with opposite  signs of energy. Therefore, such a symmetry does not impose any restrictions on elastic scattering from defects as long as there is only one state with certain value of $k_y=k_{y0}$, for which $\varepsilon (k_{y0})=\varepsilon (-k_{y0})=0$. In principle, the  $T_r$ symmetry in this case could be important if the perturbation does not break this symmetry.
However, considering the scattering matrix $\hat{S}$, which relates scattering states in different channels for the case of functions $\psi $ and $T_r\psi $ we found that $T_r$ does not impose any additional restrictions to the scattering matrix.
One should also note that the usual impurity with potential $V({\bf r})$ located at some point in the DW breaks explicitly the symmetry $T_r$ because the impurity potential is even with respect to $x\to -x$. Correspondingly, such a symmetry can not protect the edge states against scattering between ${\bf k}$ and ${\bf -k}$. Note that this
not a backward scattering since the electron velocity has the same sign for these states.

A mechanism which can lead to the absence of intravalley backward scattering is related to the peculiarity of electron band structure.
Let us consider the wave functions of electron in the edge states $|k_y\rangle =e^{ik_yy}\, \psi _{k_y}(x)$ and $|k'_y\rangle =e^{ik'_yy}\, \psi _{k'_y}(x)$ corresponding to energy $\varepsilon $ in the gap. These states belong to different bands in the same valley (see Fig.~3).
Matrix element $\langle k_y|V({\bf r})|k'_y\rangle $ of the impurity perturbation located at the point ${\bf R}=0$ (at the domain wall) is
\begin{eqnarray}
\label{18}
\langle k_y|V({\bf r})|k'_y\rangle
=\int \frac{dq_x}{2\pi }\, V(q_x,k_y-k'_y)
\nonumber \\ \times
\int _{-\infty }^\infty dx\, e^{iq_xx}\, \psi ^\dag _{k_y}(x)\, \psi _{k'_y}(x),
\end{eqnarray}
where $V(q_x,q_y)$ is the Fourier transform of the impurity potential. We calculated the wavefunctions $\psi _{k_y}(x)$ and $\psi _{k'_y}(x)$ numerically, and using these results we found that $\psi _{k_y}(x)$ and $\psi _{k'_y}(x)$ are numerically orthogonal (the integral of non-orthogonality was negligible
for the normalized wavefunctions). From this we conclude  that the probability of impurity scattering from $k_y$ to $k'_y$ is negligible.


%

\end{document}